\documentclass[final,superscriptaddress,english,twocolumn,amssymb, nobibnotes, aps, prl, longbibliography]{revtex4-1}
\usepackage{graphicx}
\usepackage{natbib}
\newcommand{\tr}{\mathrm{Tr}}
\usepackage{amsmath}

\usepackage{amssymb}
\usepackage{appendix}
\usepackage[dvipsnames]{xcolor}
\usepackage[section]{placeins}
\definecolor{darkblue}{rgb}{0,0,.65}
\definecolor{darkgreen}{rgb}{0.28,0.41,0.19}

\usepackage{wrapfig}

\usepackage[%
    pdfauthor={David J. Luitz},%
  pdfstartview=FitH,%
  breaklinks=true,%
  bookmarks=true,%
  colorlinks=true,%
  anchorcolor=black,%
  citecolor=blue,
  filecolor=black,%
  menucolor=black,%
  urlcolor=darkblue,%
  linkcolor=blue,%
 ]{hyperref}
\usepackage[all]{hypcap} %

\newcommand{\mat}[1]{\mathsf{#1}}

\newcommand{\E}{\mathrm{e}}

\newcommand{\eye}{\mathbf{1}}

\graphicspath{{images/}}
\begin{document}

\title{Hierarchy of relaxation timescales in local random Liouvillians}

\author{Kevin Wang}
\author{Francesco Piazza}
\email{piazza@pks.mpg.de}
\author{David J. Luitz}
\email{dluitz@pks.mpg.de}
\affiliation{Max Planck Institute for the Physics of Complex Systems, Noethnitzer Str. 38, Dresden, Germany}
\date{\today}

\begin{abstract}
To characterize the generic behavior of open quantum systems, we consider random, purely dissipative Liouvillians with a notion of locality. We find that the positivity of the map implies a sharp separation of the relaxation timescales according to the locality of observables.
Specifically, we analyze a spin-$1/2$ system of size $\ell$ with up to $n$-body Lindblad operators, which are $n$-local in the complexity-theory sense. 
Without locality ($n=\ell$), the complex Liouvillian spectrum densely covers a ``lemon''-shaped support, in agreement with recent findings [\href{https://journals.aps.org/prl/abstract/10.1103/PhysRevLett.123.140403}{\textit{Phys. Rev. Lett.} \textbf{123}, 140403}; \href{https://arxiv.org/abs/1905.02155}{arXiv:1905.02155}].
However, for \emph{local Liouvillians} ($n<\ell$), we find that the spectrum is composed of several dense clusters with random matrix spacing statistics, each featuring a lemon-shaped support wherein all eigenvectors correspond to $n$-body decay modes. 
This implies a hierarchy of relaxation timescales of $n$-body observables, which we verify to be robust in the thermodynamic limit.
\end{abstract}
\maketitle

\textbf{Introduction.}
For unitary quantum many-body dynamics, the characterization of generic features common to the vast majority of systems is well developed, in the form of an effective random matrix theory \cite{mehta_random_2004,wigner_characteristic_1955,dyson1_1962,dyson2_1962,dyson3_1962,bohigas_1984}. It is for example
a crucial ingredient to our understanding of thermalization in unitary quantum systems, manifest in the eigenstate thermalization hypothesis (ETH) \cite{deutsch_quantum_1991,srednicki_chaos_1994,srednicki_thermal_1996,rigol_thermalization_2008,deutsch_eigenstate_2018,dalessio_quantum_2016,borgonovi_quantum_2016}.

For open quantum many-body systems analogous organizing principles are yet missing.
Only very recently the first developments in this direction appeared with the investigation of spectral features of a purely random Liouvillian \cite{denisov_universal_2019,sa_spectral_2019,can_random_2019,can_spectral_2019,xu_extreme_2019}, 
describing generic properties of trace preserving positive quantum maps.
The purely random Liouvillians considered in \cite{denisov_universal_2019,sa_spectral_2019,can_spectral_2019} constitute the least structured models aiming at describing generic features of open quantum many-body systems. A main result of these recent studies is that the spectrum of purely random Liouvillians is densely covering a lemon-shaped support whose form is universal, differentiating random Liouvillians from completely random Ginibre matrices with circular spectrum \cite{ginibre_statistical_1965,mehta_random_2004,girko_1983,tao_random_2008}.

\begin{figure}[h]
	\centering
	\includegraphics[width=\columnwidth]{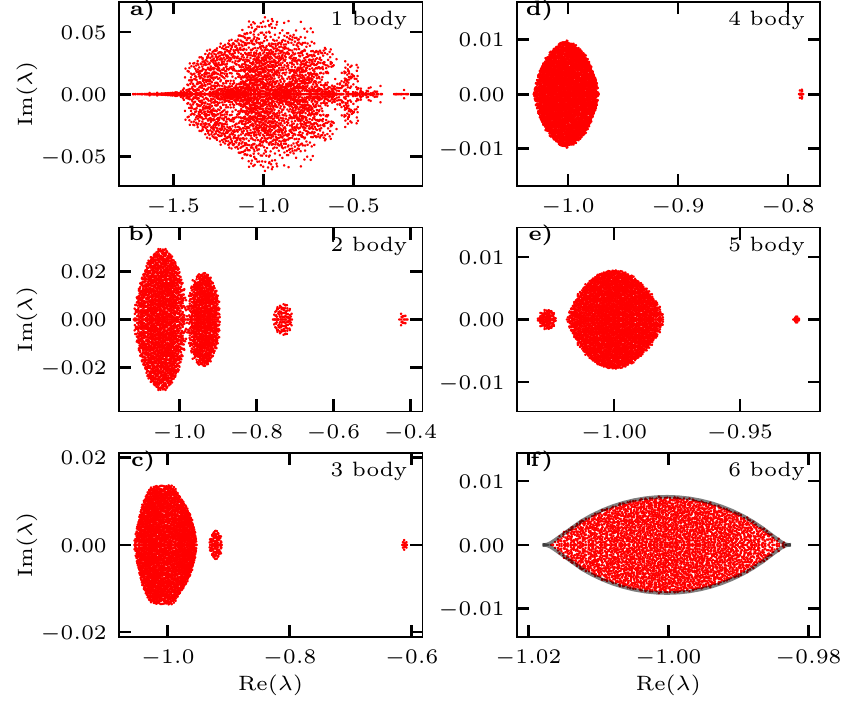}
	\caption{Eigenvalue distribution (steady state $\lambda=0$ omitted) for system size of $\ell=6$ ($N=2^\ell=64$) and different $n$ locality of the Lindblad operators $L_i$. a) $L_i$ are only one-body Pauli strings. b) Two-body Pauli strings are added, leading to four eigenvalue clusters. c)-e) Higher order Pauli strings are progressively inserted and the eigenvalue clusters begin to merge until f) the full basis of traceless matrices $\{L_n\}$ is obtained once six-body Pauli strings are included. The support of the single eigenvalue cluster matches the result for purely random Liouvillians (gray curve) \cite{denisov_universal_2019}. }
	\label{fig:L6_progression}
\end{figure}

How close is a completely random Liouvillian to a physical system? In this Letter, we make a step towards answering this question by adding the minimal amount of structure to a purely random Liouvillian, namely a notion of locality.
We consider a dissipative spin system of size $\ell$ 
with Lindblad operators given by Pauli strings classified by their length $n\leq\ell$, the latter being a measure of their $n$-locality in a complexity-theory sense \cite{kitaev_classical_2002}.

In this language, purely random Liouvillians are completely nonlocal and structureless, since they include the full operator basis including Pauli strings of all lengths $1\leq n\leq \ell$.

Here, we restrict the maximal number of non-identity operators in the Lindblad operators to $n_{\rm max}$ and examine the spectral properties of the adjoint Liouvillian as $n_{\text{max}}$ is varied, $n_{\text{max}}=\ell$ corresponding to the nonlocal case.

\begin{figure*}
	\centering
	\includegraphics[width=\textwidth]{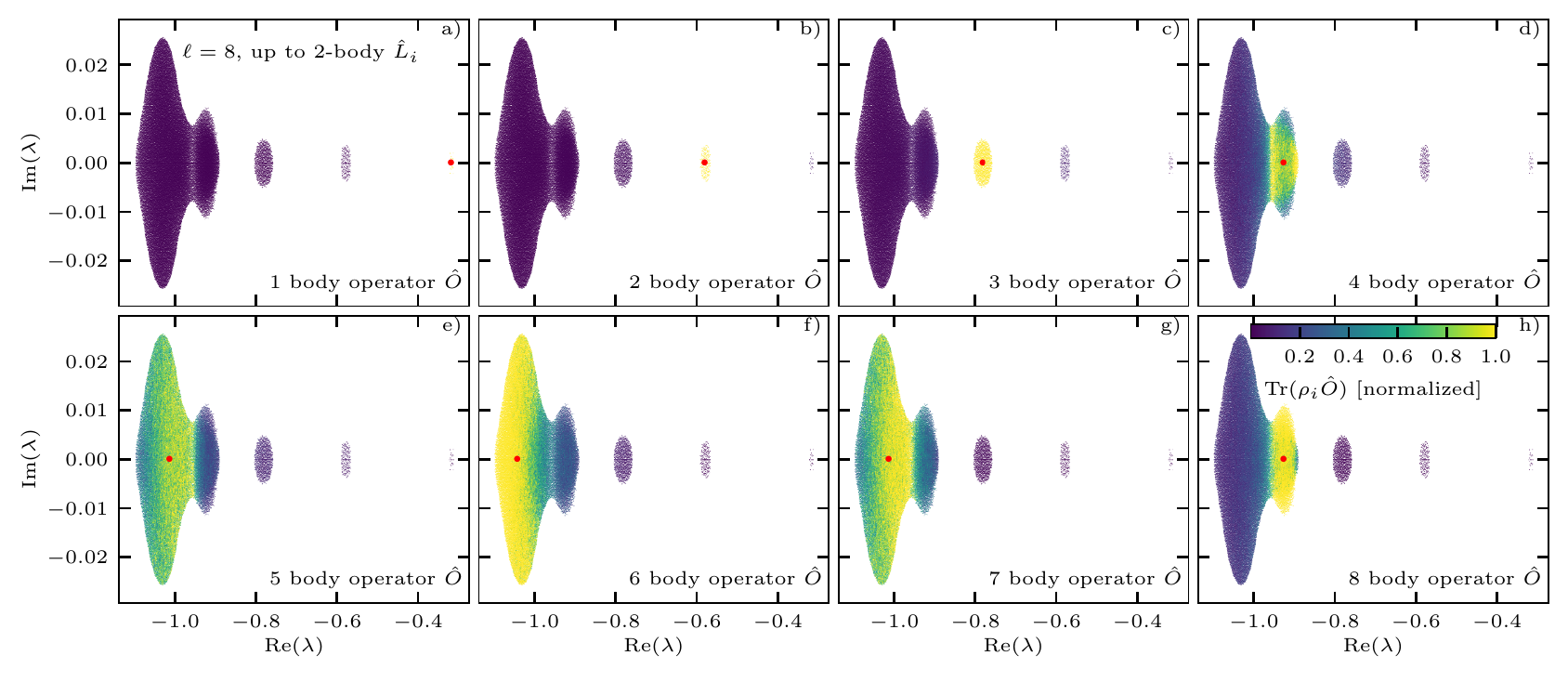}
	\caption{Color: expectation value $\tr( \rho_i O )$ of $n$-local operators $O$ given by random superpositions of $n$-body Pauli strings for all (except steady state) eigenvalues $\lambda_i$ of the 2-local Liouvillian (i.e. $n_\text{max}=2$) with the corresponding decay modes $\rho_i$, $\ell=8$. The coloring of the eigenvalue markers is done for random one-body $O$ a), two body $O$ b) etc.
	It is clearly  visible that the eigenvalue clusters correspond to the $n$ locality of the decay modes, revealing different relaxation timescales based on the degree of locality of the observable $O$. The red dot indicates the analytical prediction of the centers of $n$ local eigenvalue clusters (cf. main text).
	\label{fig:L8_highlighting}}
\end{figure*}

We find that the complex spectrum generically decomposes into several simply-connected, lemon-shaped eigenvalue clusters, whose center of symmetry lies on the real axis (cf. Fig.~\ref{fig:L6_progression} ). Each well isolated cluster corresponds to decay modes (right eigenvectors of $\mathcal{L}$) which govern the relaxation of $n$-local observables. In other words, the eigenvalue clusters are organized by the complexity of observables, whose relaxation their eigenvectors control (cf. Fig.~\ref{fig:L8_highlighting}).

\begin{figure}[t]
	\centering
	\includegraphics[width=\columnwidth]{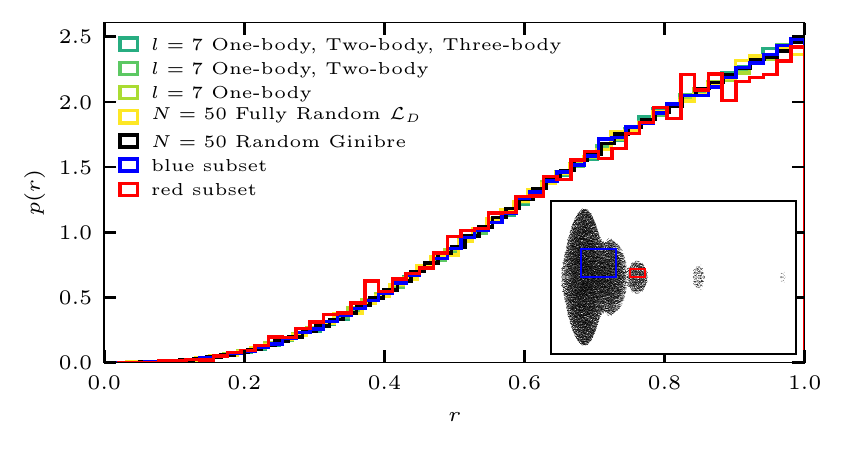}
	\caption{Probability density of the spectral gap ratio $r = |\lambda_0-\lambda_1|/|\lambda_0-\lambda_2|$ computed for each complex eigenvalue $\lambda_0$ and its nearest ($\lambda_1$) and next nearest neighbor ($\lambda_2$). We compare the eigenvalues of Liouvillians of a system of size $\ell=7$ ($100$ realizations) of different $n$-locality to random Ginibre ensembles (1000 realizations) and completely random ($n=\ell$) Liouvillians. Due to the symmetry of the spectrum across the real axis for random Liouvillians -- which disproportionately weights ratios of 1 near the real line when next-nearest and nearest neighbor distances are the same -- only the eigenvalues for which Im($\lambda)>0$ are included in the statistics. We also show results for completely random dissipative Liouvillian (full operator basis, 100 realizations, yellow) as in \cite{denisov_universal_2019} and analyze two subsets of eigenvalues (inside the blue/red rectangles in the inset). }
	\label{fig:ratio_hist}
\end{figure}

If two clusters overlap (typically for large $n$), they  merge as opposed to just joining their supports as their respective eigenvalues interact and exhibit level repulsion (cf. Fig. \ref{fig:ratio_hist}), such that the merged cluster exhibits random matrix statistics.

\begin{figure}
    \centering
    \includegraphics[width=\columnwidth]{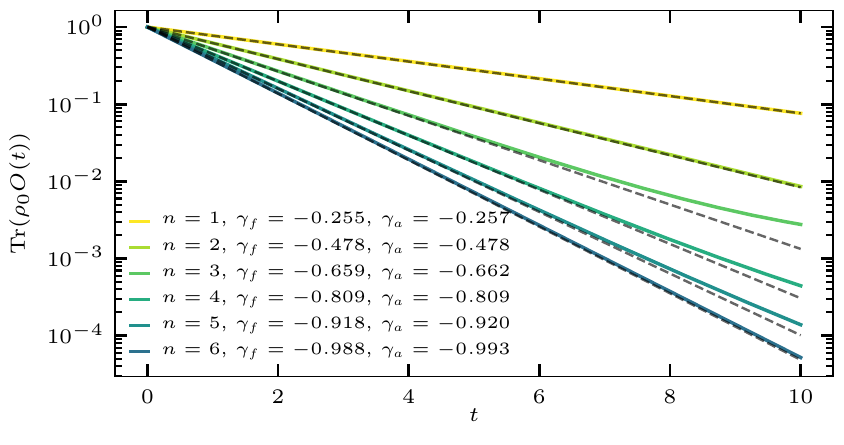}
    \caption{Numerical calculations of operator decay of different $n$-body observables vs. analytic predictions (dashed) (cf. main text) for a system of size $\ell=10$ and $n_\text{max}=2$. A slight discrepancy between the predicted ($\gamma_a$) and fitted ($\gamma_f$) decay rates is expected due to finite width of the eigenvalue clusters.}
    \label{fig:L10_relaxation}
\end{figure}

\begin{figure}
    \centering
    \includegraphics[width=\columnwidth]{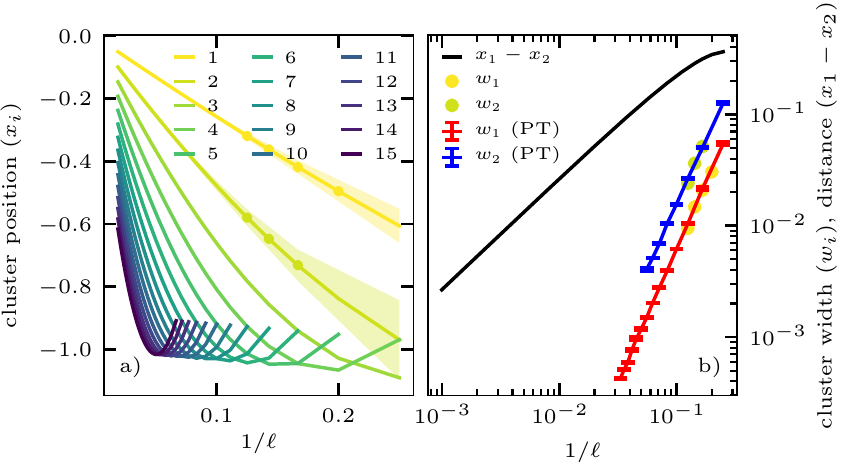}
    \caption{Left: Analytical predictions of eigenvalue cluster centers $x_i$ vs. $1/\ell$, shaded area indicates the cluster width $w_i=\text{max}[ \text{Re}(\lambda_i)]-\text{min}[ \text{Re}(\lambda_i) ]$ from degenerate perturbation theory (PT), circles are exact positions from full diagonalization of $\mathcal L$. Right: Comparison of the distance $x_1-x_2$ of the first two clusters with their width $w_i$. Circles correspond to estimates of the width from full diagonalization of $\mathcal L$, the red/blue points with errorbars correspond to degenerate perturbation theory up to $\ell=30$, averaged over $\approx 20$ realizations.}
    \label{fig:lemon_center_progression}
  \end{figure}

Our findings correspond to a separation of relaxation-timescales between observables with $n$-locality, e.g. one-body operators decay at a different rate compared to two-body operators etc. (cf. Fig.~\ref{fig:L10_relaxation}). This hierarchy, which we characterize analytically and confirm by time-evolution, is present for sufficiently local Liouvillians and observables, and persists in the thermodynamic limit, since the separation of the eigenvalue clusters decays more slowly than their width (cf. Fig.~\ref{fig:lemon_center_progression}). 
The source of this hierarchy is in the positivity of the map and is therefore special to Liouvillian dynamics, while absent in the Hamiltonian counterpart.

\textbf{Local Liouvillian model.}
The simplest description of the dynamics of a markovian open quantum system is in terms of a quantum master equation $\partial_t \rho_t = \mathcal{L}(\rho_t)$, where $\mathcal L$ is a Liouvillian superoperator.

A generic Liouvillian takes the Gorini-Kossakowski-Sudarshan-Lindblad form \cite{breuer2002theory,gorini_kossakowski_76,Lindblad1976}
$\mathcal{L}(\rho)=\mathcal{L}_H(\rho) + \mathcal{L}_D(\rho)$,
where $\mathcal{L}_H(\rho)=-i[H,\rho]$ is the unitary evolution with an hermitian Hamiltonian $H$  and 
\begin{equation}
\mathcal{L}_D(\rho)=\sum_{i,j=1}^{N_L}K_{ij}[L_i\rho L_j^\dagger - \frac{1}{2}\{L_j^\dagger L_i,\rho\}]
\end{equation}
\noindent is the non-unitary contribution, where the braces denote the anti-commutator. 
The traceless Lindblad operators $L_i$, ${i=1,2,\ldots,N}$, ($N_L\leq N=\mathrm{dim}\mathcal{H}$ is the size of the Hilbert space), satisfy the orthonormality condition Tr$(L_iL_j^\dagger)=\delta_{ij}$ and constitute a complete basis of the associated operator space.
If the Kossakovski matrix $K$ is positive semidefinite, the generated time-evolution of the density matrix $\rho$ preserves the trace and positivity of $\rho$. 
Here we restrict the discussion to purely non-unitary evolution $\mathcal{L}_H=0$, leaving the effect of the Hamiltonian for future studies and thus henceforth we use $\mathcal{L}=\mathcal{L}_D$. An alternative approach has been considered by studying ergodic many-body Hamiltonians perturbed by a non-hermitian term \cite{luitz_EPs_2019,Heiss_1990,Birchall_2012,cejnar_2018}.

We focus on dissipative quantum spin systems of $\ell$ spins $1/2$, with a tensor product Hilbert space  $\bigotimes\limits_{i=1}^\ell \mathcal{H}_i$. 

To introduce a notion of locality in our system, we use 
Lindblad operators $L_i$ given by normalized Pauli strings $S_{\vec{x}} = 1/\sqrt{N} \sigma_{x_1} \otimes \sigma_{x_2}\dots \otimes\sigma_{x_\ell}$ ($x_i\in{0,1,2,3}$), and classify them by the number $n=\sum_i (1-\delta_{x_i,0})$ of non-identity Pauli matrices (such that they correspond to $n$-body operators, and are $n$-local \cite{kitaev_classical_2002}), including terms up to $n\leq n_\text{max}$.

$N_n=\binom{\ell}{n}\left((\mathrm{dim}\mathcal{H}_i)^2-1\right)^n$ is the dimension of the space of all $n$-local operators.
For $n_{\rm max}=\ell$ the total number $N_L=\sum_{n=1}^{n_{\rm max}}N_n$ of traceless operators employed becomes equal to the size of a complete basis $N^2-1$ with $N=(\mathrm{dim}\mathcal{H}_i)^\ell$. For our spin-$1/2$ system, $N=2^\ell$.

In the present work, we are interested in the \emph{generic} features of dissipative $n$-local Liouvillians, and therefore consider \emph{random} Kossakowski matrices. Each $K$ matrix is generated by sampling its (non-negative) eigenvalues $d_i$ from a box distribution and then transforming it to a random basis with a random unitary $U$ sampled from the Haar measure, $K=U^\dagger D U$ is normalized by $\tr K = N$.

Because $K$ is positive semidefinite, the mean of its eigenvalues is positive $d=\text{mean}(d_i)=\tr K/N_L > 0$.

\textbf{Spectrum of local Liouvillians.}
In Fig. \ref{fig:L6_progression}, we show the spectrum of the Liouvillian $\mathcal{L}$ for one random realization of $K$ and different degrees of locality $n_\text{max}$ of the involved Lindblad operators. For $n_\text{max}=1$, $L_i$ act nontrivially only on a single site and the spectrum shown in Fig. \ref{fig:L6_progression}a) shows several smeared-out eigenvalue clusters. Once two body operators are added: $n_\text{max}=2$, the spectrum shows clearly distinct eigenvalue clusters (cf. Fig \ref{fig:L6_progression}b), each of which is composed of a different number of eigenvalues. Further increasing $n_{\rm max}$ leads to a merging of the eigenvalue clusters, and once all possible traceless operators are included $n_{\rm max}=\ell$ (thus removing the locality entirely), a single eigenvalue cluster remains, which has the shape of a ``lemon''. In Ref. \cite{denisov_universal_2019}, an analytical form of the envelope of spectra of completely random Liouvillians was derived, which is shown as the gray curve in Fig. \ref{fig:L6_progression}f).
We notice that the individual shapes of the different eigenvalue clusters we obtain are close to this lemon shape. Furthermore, different realizations of the random $K$ matrix yield very similar results.

Before we discuss the origin of the separation of eigenvalue clusters, we analyze the statistics of complex level spacings. Complex spectra have no ordering, but it is straightforward to generalize level spacing ratios \cite{oganesyan_localization_2007} (which get rid of the otherwise necessary unfolding of the local density of states) to the complex case by finding the nearest $\lambda_1$ and next nearest neighbor $\lambda_2$ of an eigenvalue $\lambda_0$ and defining $r = |\lambda_0-\lambda_1|/|\lambda_0-\lambda_2|$. This definition was recently generalized to include phase angles \cite{sa_complex_2019}.

In Fig. \ref{fig:ratio_hist}, we show our results for the distribution of level spacing ratios $r$ for a system of size $\ell=7$ and different $n$-locality of the included Lindblad operators. We compare the distribution to the case of completely random Liouvillians as studied in Refs. \cite{denisov_universal_2019,sa_spectral_2019} and to random square matrices with gaussian matrix elements (i.e. from the Ginibre ensemble). All spacing distributions are identical and it does not matter whether we consider all eigenvalues or subsets (as indicated in the inset). This confirms that local spectral properties of generic Liouvillians admit a random matrix description and demonstrates that eigenvalue clusters which have merged (cf. blue rectangle subset) do not simply overlap but also show level repulsion.

\textbf{Hierarchy of relaxation timescales.}
The existence of distinct clusters with different real parts of the corresponding eigenvalues suggests that there are different relaxation timescales in the system. The corresponding eigenvectors are decay modes and we analyze their properties in the following. Since the Liouvillian contains $n$-local terms, it is natural to consider the relaxation of $n$ local operators $O$. The time evolution of such an operator is given by $\tr(\rho(t){O} )$ and each decay mode $\rho_m$ contributes with a term $\E^{\lambda_m} \tr(\rho_m {O})$, where $\lambda_m$ is the corresponding eigenvalue of $\mathcal L$ and hence $\text{Re}(\lambda_m)$ sets the relaxation timescale of this term. It is natural to ask which relaxation modes contribute most to each locality class of operators $O$. We show in Fig. \ref{fig:L8_highlighting} for a system of size $\ell=8$ with a $2$-local Liouvillian (i.e. $n_\text{max}=2$) all eigenvalues $\lambda_i$ of the Liouvillian (except the steady state $\lambda_0=0$) for one realization of a random Kossakowski matrix $K$. Each eigenvalue is shown in a color corresponding to the contribution of its eigenstate (decay mode) $\rho_i$ to a \emph{random} $k$-local operator $O^{(k)}$, given by {$\tr(\rho_i O^{(k)})$}. Panel \ref{fig:L8_highlighting}a) shows eigenvalue coloring for the relaxation of a random one body operator, panel b) for a two body operator, panel c) for a three body operator etc.
It is obvious from this analysis that only eigenvalues from the cluster with the largest real part contribute to the relaxation of one-body observables (panel a), the second eigenvalue cluster corresponds to two body observables and the third to three body observables. 
More complex observables fall in the largest eigenvalue cluster which has an interesting substructure: its left part seems to include 5, 6, and 7 body observable relaxation, its right part covers 4 and 8 body operators.

This finding thus implies a separation between the timescales of relaxation of observables with a different degree of locality. We confirm the existence of such a hierarchy of relaxation timescales in a dynamical simulation, where we directly study the equilibration of random $n$-body operators in Fig. \ref{fig:L10_relaxation} for a system of size $\ell=10$.

\textbf{Perturbation theory.}
To understand the rich spectral structure observed in Fig. \ref{fig:L8_highlighting}, it is helpful to consider the \emph{adjoint Liouvillian} describing the evolution of an operator $X$
\begin{equation}
\mathcal{L}_a[X] =  \sum_{n,m}^{{N_L}} K_{n,m} (L_m^\dagger X L_n - \frac 1 2 \{ L_m^\dagger L_n, X \}),
\end{equation} 
keeping in mind that the Lindblad operators $L_n =  S_{\vec{y}_n}$ are given by normalized Pauli strings. 
Since the matrix $K$ is sampled in its diagonal basis as $D$ and rotated to a random basis $K=U^\dagger D U$ with a CUE matrix $U$, it is easy to show that $\text{mean}(K_{nn}) = \frac{N}{N_L} = d$, while the mean of the offdiagonal matrix elements of $K$ vanishes. The standard deviation of its matrix elements is given by $\text{std}(K_{nm})=N/(\sqrt{6}N_L^{3/2})$. Therefore $K$ is typically \emph{diagonally dominant}.

Thus, we can consider the offdiagonal part of $K$ as a perturbation and decompose $K=d \eye + K'$. Then, we represent $\mathcal{L}_a$ in the basis of normalized Pauli strings with matrix elements $\mat{L}_{xy} = \tr\left(S_{\vec{x}} \mathcal{L}_a [S_{\vec{y}} ] \right)$. We define $\mat L = \mat L^0 + \mat L^1$, where $\mat L^0$ contains all terms stemming from the constant part $d\eye$ of $K$ and $\mat L^1$ contains the rest $K'$.

It turns out that the offdiagonal parts $(\vec{x}\neq \vec{y})$ of $\mat{L}^0$ vanish, while the diagonal is given by
\begin{equation}
\mat{L}_{xx}^0 = d \sum_{n=1}^{N_L} \left[ \tr\left( S_{\vec{x}} S_{\vec{y}_n} S_{\vec{x}} S_{\vec{y}_n} \right) - \frac{1}{N} \tr\left(S_{\vec{x}}^2\right) \right].
\label{eq:liou_diag}
\end{equation}
The terms in the sum in \eqref{eq:liou_diag} vanish if $[S_{\vec{x}},S_{\vec{y}_n}]=0$, otherwise, they yield $-2/N$ if the number of differing Pauli matrices on the sites where both $S_{\vec{x} }$ and $L_{n}$  are not identity is odd, and yield zero if this number is even. Counting the nonzero terms is a simple matter of combinatorics, the result depends only on the number $m$ of nonidentity Pauli matrices in the $L_{n}$ operators, the number $n_x$ of nonidentity Pauli matrices in $S_{\vec{x}}$ and $\ell$: then $f_m(n_x,\ell)$ is the number of noncommuting terms. For $m=1$, we get $f_1(n_x,\ell)=2 n$, for $m=2$, $f_2(n_x,\ell) = 6n_x\ell - 4n_x^2-2n_x$.

In total, the eigenvalues of $\mat L^0$ are therefore 
\begin{equation}
\mat{L}^0_{xx} = \frac{-2}{N_L} \sum_{m=1}^{n_\text{max}} f_m(n_x,\ell),
\end{equation}
and depend (for a fixed set of $n_\text{max}$-local $L_n$) \emph{only} on the $n_x$-locality of the basis operator $S_{\vec{x}}$, making them highly degenerate. They represent the leading-order perturbation theory result and we show the corresponding values as red dots in Fig. \ref{fig:L8_highlighting}, where they perfectly predict the location of the $n_x$-local eigenvalue clusters (see also Fig.~\ref{fig:lemon_center_progression}).

The last step is to use nonhermitian degenerate perturbation theory to lift this degeneracy and to calculate the spread of the eigenvalue clusters. To do this, we simply diagonalize $\mat L^1$ in the degenerate block of $n$-local operators. We summarize our results in Fig. \ref{fig:lemon_center_progression}, where we show in a) the analytical result for the centers of the $n$-local ``lemons'' together with the numerical results, showing a perfect match.  Panel \ref{fig:lemon_center_progression}b) compares the width of the first two ($1$-local and $2$-local) clusters to their distance, showing that the clusters remain well separated in the thermodynamic limit, since the width of the clusters decays much faster compared to their distance. The validity of perturbation theory is confirmed by the perfect match of the predicted width compared to the full calculation (up to $\ell=8$).

\textbf{Conclusion.}
We have shown that, for random local Liouvillians, observables with different degrees of locality relax with sharply separated timescales. This hierarchy, while being generic for local Liouvillians, is not present in the Hamiltonian counterpart.

\bibliography{louvillian-level-stats.bib, louvillian-level-stats-auxiliary.bib}

\end{document}